\documentclass{article}
\usepackage{spconf,amsmath,amssymb,graphicx,cite,hyperref}
\usepackage{algorithm,algpseudocode}
\usepackage{tikz}
\usepackage[table]{xcolor}
\definecolor{surablue}{RGB}{235,245,255}
\hypersetup{hidelinks,hypertexnames=false}
\usepackage{multirow, multicol}

\newcommand{\circnum}[1]{%
  \tikz[baseline=(char.base)]{
    \node[
      shape=circle,
      fill=black,
      text=white,
      font=\scriptsize\bfseries,
      inner sep=0.4pt
    ] (char) {#1};
  }%
}

\newcommand{\contribbullet}{%
  \raisebox{0.05ex}{\large$\bullet$}\hspace{0.3em}%
}

\makeatletter
\long\def\@makecaption#1#2{\vskip 4pt
 \setbox\@tempboxa\hbox{#1. #2}
 \ifdim \wd\@tempboxa >\hsize #1. #2\par \else \hbox
 to\hsize{\hfil\box\@tempboxa\hfil}\fi}
\makeatother

\title{Data-Efficient Unlearning-Based Backdoor Detection in Federated Learning}

\name{Dowon Kim, Junhao Cai, Sung Il Choi, Dohun Kim, Juhyun Park, Chengjun Jin, Changhee Joo$^{*}$
\thanks{$^{*}$Corresponding author: Changhee Joo (\texttt{changhee@korea.ac.kr}).}}

\address{Department of Computer Science and Engineering \\
         Korea University, Republic of Korea}
\makeatletter
\newcommand{\ps@arxivpreprint}{%
  \ps@empty
  \def\@oddfoot{\parbox[b]{\textwidth}{\footnotesize\raggedright
    This work has been submitted to the IEEE for possible publication.\newline
    Copyright may be transferred without notice, after which this version
    may no longer be accessible.}}%
  \let\@evenfoot\@oddfoot
}
\makeatother

\begin{document}
\ninept
\maketitle
\thispagestyle{arxivpreprint}
\begin{abstract}
Detecting backdoor attacks in federated learning (FL) is challenging
because malicious clients can manipulate the global model while
preserving main-task performance. Static model-update statistics may
therefore fail to clearly distinguish malicious from benign clients.
Unlearning-based detectors instead probe client models through loss
ascent on clean server data, but detection performance degrades with
scarce proxy data. We propose Screening via Unlearning Response
Analysis (SURA), a data-efficient detector. SURA reuses the same small
clean proxy across multiple train-mode loss-ascent steps, strengthening
client responses without additional clean data. For federated vision
models with batch normalization (BN), repeated probing also changes BN
running statistics and can confound endpoint measurements. SURA
addresses this by restoring the shared pre-probe BN state only at
readout while preserving the parameter changes induced by probing.
This separates repeated probing from response measurement and removes
BN drift and the interaction between parameter and BN-state changes
from the readout. Across six backdoor attacks, SURA uses only 10 clean
server images, outperforms methods under the same data budget, and
approaches the robustness obtained with much larger proxy sets.
\end{abstract}

\section{Introduction}
\label{sec:intro}

\begin{figure}[!t]
\centering
\includegraphics[width=\columnwidth]{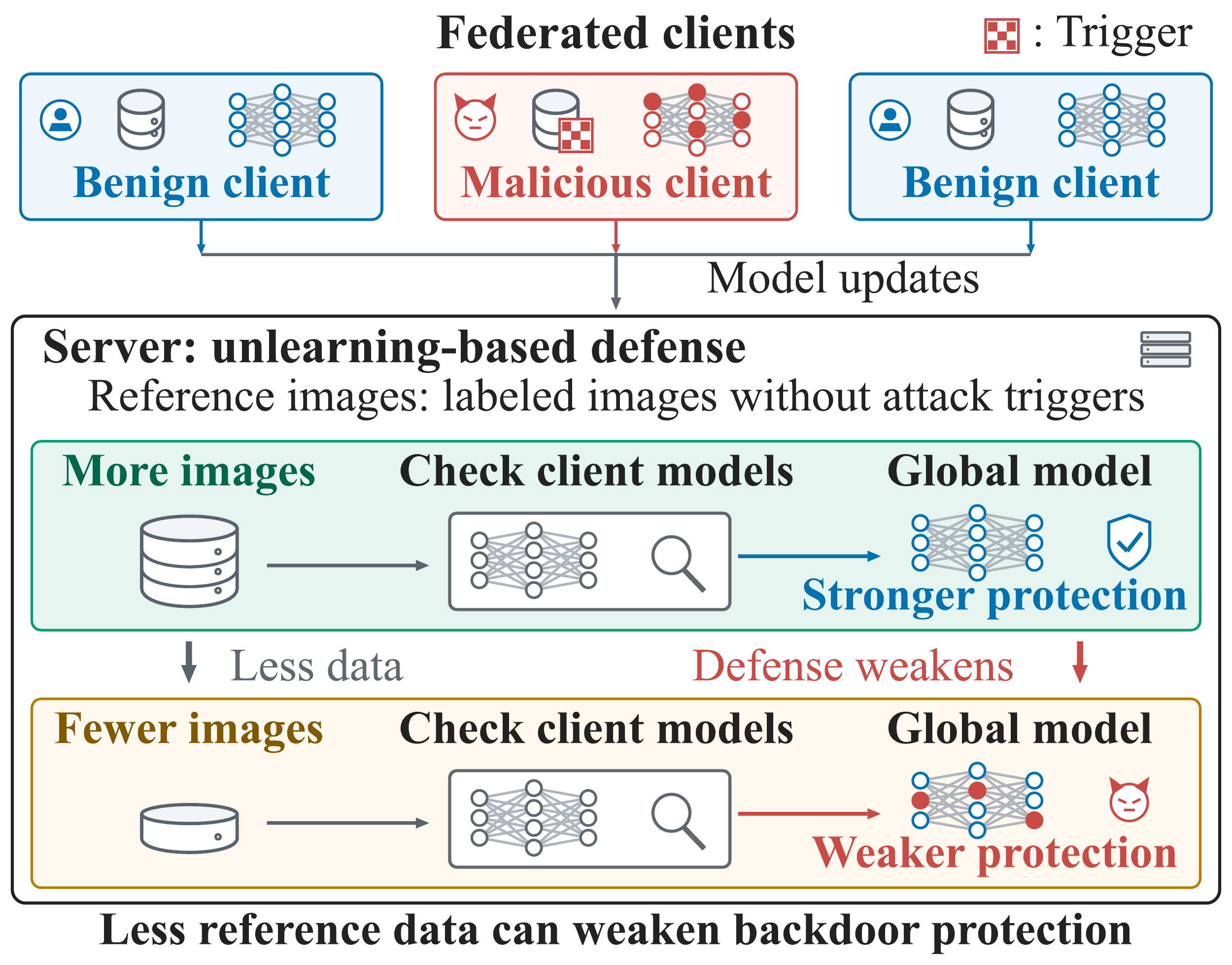}\par
\caption{Illustration of the effect of limited server reference data
on unlearning-based backdoor screening.}
\label{fig:srfp_motivation}
\end{figure}

Federated learning (FL) enables multiple clients to collaboratively train a shared model while keeping their raw data local~\cite{pmlr-v54-mcmahan17a}. However, because the server cannot directly inspect clients’ local training, malicious clients can poison their data or manipulate model updates to implant backdoors~\cite{pmlr-v108-bagdasaryan20a}. Backdoor attacks are particularly difficult to detect because they preserve performance on benign inputs while inducing attacker-chosen behavior on inputs containing an attack trigger pattern~\cite{8685687}. Existing server-side defenses commonly examine submitted updates using distances, similarities, clustering, or robust aggregation~\cite{NIPS2017_f4b9ec30,259745,280048,9721118}. However, because backdoored updates can remain statistically similar to benign ones, such update-based signals may provide limited separation between malicious and benign clients~\cite{Huang_2023_ICCV}. To overcome this limitation, we exploit unlearning-based detection that actively probes each client model with labeled, trigger-free server data (a clean proxy) and identifies suspicious clients from differences in their
loss responses~\cite{Xu_2025_WACV}.
Fig.~\ref{fig:srfp_motivation} illustrates this low-data setting and
its impact on unlearning-based screening.

We use probing to denote clean-loss ascent on the proxy and readout to denote the subsequent loss measurement. An endpoint readout measures loss after the final probe step. With few proxy images, shallow probing may be insufficiently discriminative, motivating deeper probing. For the vision models with batch normalization (BN) considered here, deeper train-mode probing also updates BN running means and variances~\cite{pmlr-v37-ioffe15}. As a result, the endpoint response mixes probe-induced parameter changes with BN-buffer drift and their interaction.

This leads to two core challenges.
\circnum{1} \textbf{Discriminative probing under scarce data.}
The probe must elicit sufficiently discriminative loss responses from only a handful of clean server samples without relying on a large proxy set.
\circnum{2} \textbf{Comparable response measurement.}
Client responses should be measured under a common BN reference
rather than client-specific post-probe states. Together, these challenges call for a design that strengthens probing under scarce data while  decoupling response measurement from client-specific BN drift.

To address these challenges, we propose Screening via Unlearning Response Analysis (SURA). SURA first applies repeated loss-ascent probing to elicit informative client responses. It then restores the shared pre-probe global BN statistics only when measuring those responses, while preserving the parameter changes induced by probing. This design separates the probing process from the state used for response measurement and provides a common BN reference across clients.
The separation allows deeper probing to amplify weak responses without forcing response measurement to inherit the BN state accumulated during probing. Our scope is limited to models with BN running statistics; extending the principle to architectures without stateful BN remains future work.

\textbf{Our contributions are summarized as follows:}
\contribbullet We show that endpoint responses in unlearning-based backdoor detection can mix parameter changes with BN-buffer drift and their interaction, and derive an exact four-state decomposition
of these effects.
\contribbullet We propose SURA, which repeatedly reuses a small clean
proxy to increase probing depth without additional clean data, and
separates this probing process from response readout through shared
pre-probe BN restoration.
\contribbullet We evaluate SURA across six backdoor attacks, achieving a 0.53\% mean attack success rate with only 10 clean server images on CIFAR-100/VGG16-BN and approaching the robustness obtained with 500-image proxy sets. We further study proxy sensitivity, generalization, and adaptive attacks.

\section{METHODOLOGY}
\label{sec:method}

\subsection{Defense Setting and Probe Models}
\label{ssec:threat_model}

Let $\mathcal{S}_t$ denote the participating clients at round $t$.
The global state is
$\boldsymbol{w}^t=(\boldsymbol{\theta}_t,\boldsymbol{b}_t)$, where
$\boldsymbol{\theta}_t$ stacks all trainable parameters, including BN
affine parameters, and $\boldsymbol{b}_t$ stacks all BN buffers,
including running means, running variances, and batch counters.
Each client~$i$ returns an update vector
$\boldsymbol{\delta}_i^t$ relative to the full state
$\boldsymbol{w}^t$, including both trainable parameters and BN buffers.

The server holds a small labeled, trigger-free proxy set
$\mathcal{D}_{\mathrm p}$ but has no access to private client data. An adversary controls a subset of clients and may modify their data, training, and submitted updates to implant a targeted backdoor while preserving clean accuracy. The server knows neither which clients are malicious nor the trigger or target label, and the exact server proxy is unavailable to the adversary. The defense aims to reject malicious updates while limiting false rejection of benign clients.

\begin{equation}
\widetilde{\boldsymbol{w}}_i^t
=\boldsymbol{w}^t
+\lambda\boldsymbol{\delta}_i^t
+(1-\lambda)\bar{\boldsymbol{\delta}}^t,
\label{eq:client_conditioning}
\end{equation}
where $\bar{\boldsymbol{\delta}}^t= \frac{1}{|\mathcal{S}_t|}
\sum_{j\in\mathcal{S}_t}\boldsymbol{\delta}_j^t$ is the round-average update and $\lambda\in[0,1]$ is the fusion weight.
$|\cdot|$ denotes the set cardinality. Let $(\boldsymbol{\theta}_{i,0},\boldsymbol{b}_{i,0})$ denote the initial probe state.
We initialize $\boldsymbol{\theta}_{i,0}$ using only the trainable parameters of $\widetilde{\boldsymbol{w}}_i^t$, 
and $\boldsymbol{b}_{i,0}=\boldsymbol{b}^t$
from the unfused global model,
rather than statistics estimated from the proxy.

\subsection{Repeated Clean-Loss Probing}
\label{ssec:unlearning_response}

The repeated probing stage is summarized in Algorithm~\ref{alg:srfp}.
With a small clean proxy, shallow probing may provide weak separation
between benign and malicious client responses. SURA therefore treats
probe depth as a design axis and reuses the same proxy batch for $K$
train-mode ascent steps on the mean proxy cross-entropy loss
$\mathcal{L}_{\mathrm{p}}^{\mathrm{train}}$. This increases probe depth
without requiring additional clean samples. During probing, BN uses
batch statistics and updates its running statistics, yielding the probe
state $(\boldsymbol{\theta}_{i,K},\boldsymbol{b}_{i,K})$. Deeper probing
can strengthen response discrimination but also changes the BN state,
motivating separate response measurement. We use step size $\eta$,
momentum $\mu$, and a gradient-norm limit $c_{\mathrm{g}}$.

\begin{algorithm}[!t]
\caption{Screening via Unlearning Response Analysis (SURA)}
\label{alg:srfp}
\small
\begin{algorithmic}[1]

\Statex \textbf{Input.} $\boldsymbol{w}^t$,
$\{\boldsymbol{\delta}_i^t\}_{i\in\mathcal{S}_t}$,
$\mathcal{D}_{\mathrm{p}}$

\Statex \textbf{Parameter.} $\lambda$, $K$, $\eta$, $\mu$,
$c_{\mathrm{g}}$, $\tau_{\mathrm{SR}}$

\Statex \textbf{Output.} Accepted clients $\mathcal{B}_t$,
updated state $\boldsymbol{w}^{t+1}$

\For{each client $i\in\mathcal{S}_t$}
  \State Form $\widetilde{\boldsymbol{w}}_i^t$
         using Eq.~\eqref{eq:client_conditioning}
  \State $(\boldsymbol{\theta},\boldsymbol{b},\boldsymbol{h})\gets
         (\operatorname{params}(\widetilde{\boldsymbol{w}}_i^t),
         \boldsymbol{b}^t,\boldsymbol{0})$
  \State $\ell_0\gets
         \mathcal{L}_{\mathrm{p}}^{\mathrm{eval}}
         (\boldsymbol{\theta},\boldsymbol{b})$

  \For{$k=1,\ldots,K$}
    \State $\boldsymbol{h}\gets
           \mu\boldsymbol{h}+
           \operatorname{clip}_{c_{\mathrm{g}}}
           \!\left(
           \nabla_{\boldsymbol{\theta}}
           \mathcal{L}_{\mathrm{p}}^{\mathrm{train}}
           (\boldsymbol{\theta},\boldsymbol{b})
           \right)$
           \Comment{BN buffers update in train mode}
    \State $\boldsymbol{\theta}\gets
           \boldsymbol{\theta}+\eta\boldsymbol{h}$
  \EndFor

  \State $\boldsymbol{b}\gets\boldsymbol{b}^t$
         \Comment{Restore BN buffers}
  \State $v_i^t\gets
         \mathcal{L}_{\mathrm{p}}^{\mathrm{eval}}
         (\boldsymbol{\theta},\boldsymbol{b})-\ell_0$
\EndFor

\State Compute $\{z_i^t\}_{i\in\mathcal{S}_t}$
       using Eq.~\eqref{eq:srfp_score}
\State $\mathcal{B}_t\gets
       \{i\in\mathcal{S}_t\mid
       z_i^t\leq\tau_{\mathrm{SR}}\}$
\State Aggregate original updates using Eq.~\eqref{eq:aggregation}

\end{algorithmic}
\end{algorithm}

\begin{table*}[t]
\centering
\caption{Performance comparison on CIFAR-100 with VGG16-BN.
Attack entries report ASR (\%), and Mean averages the six attacks
unless otherwise noted. Bold and underline denote the best and
second-best results within the at-most-10-image defense block.}
\label{tab:main_results_vgg}
\small
\setlength{\tabcolsep}{1.2pt}

\begin{tabular}{lccccccccc}
\hline
\multirow{2}{*}{Method}
& \multirow{2}{*}{Images}
& \multirow{2}{*}{MA$\uparrow$}
& \multicolumn{6}{c}{Attacks (ASR \%)$\downarrow$}
& \multirow{2}{*}{Mean ASR$\downarrow$} \\ \cline{4-9}

& & & BadNets & DBA & Neurotoxin & Scaling & PGD & LIE & \\
\hline

FedAvg & -- & 53.27
& $99.69{\pm}0.03$
& $99.20{\pm}0.79$
& $97.70{\pm}1.39$
& $99.90{\pm}0.05$
& $99.76{\pm}0.12$
& $99.68{\pm}0.13$
& 99.32 \\

FedAvg (oracle) & -- & 52.71
& $0.29{\pm}0.19$
& $0.29{\pm}0.19$
& $0.29{\pm}0.19$
& $0.29{\pm}0.19$
& $0.29{\pm}0.19$
& $0.29{\pm}0.19$
& 0.29 \\

\hline
\multicolumn{10}{l}{\textit{Larger server-data references}} \\

MASA-P500~\cite{Xu_2025_WACV} & 500 & 52.52
& $0.33{\pm}0.19$
& $0.36{\pm}0.26$
& $0.38{\pm}0.24$
& $0.32{\pm}0.22$
& $0.37{\pm}0.25$
& $0.35{\pm}0.20$
& 0.35 \\

FLTrust~\cite{DBLP:conf/ndss/CaoF0G21} & 500 & 27.12
& $97.43{\pm}1.19$
& $95.32{\pm}3.09$
& $73.44{\pm}11.56$
& $97.43{\pm}1.19$
& $97.10{\pm}0.92$
& $96.23{\pm}1.86$
& 92.83 \\

\hline
\multicolumn{10}{l}{\textit{Defenses using at most 10 server images}} \\

RFA~\cite{9721118} & 0 & 49.53
& $99.39{\pm}0.07$
& $98.10{\pm}0.79$
& $95.00{\pm}2.34$
& $53.32{\pm}27.59$
& $99.23{\pm}0.22$
& $99.48{\pm}0.18$
& 90.75 \\

Multi-Krum~\cite{NIPS2017_f4b9ec30} & 0 & 43.59
& $99.07{\pm}0.66$
& $98.62{\pm}0.65$
& $93.58{\pm}4.33$
& $0.44{\pm}0.09$
& $98.91{\pm}0.85$
& $99.20{\pm}0.64$
& 81.64 \\

FoolsGold~\cite{259745} & 0 & 37.05
& $62.76{\pm}44.27$
& $61.85{\pm}43.71$
& $11.43{\pm}12.80$
& $66.56{\pm}46.95$
& $63.13{\pm}44.12$
& $37.73{\pm}41.58$
& 50.58 \\

Multi-Metrics~\cite{Huang_2023_ICCV} & 0 & 33.36
& $99.89{\pm}0.10$
& $99.95{\pm}0.04$
& $99.85{\pm}0.12$
& $67.27{\pm}46.15$
& $99.91{\pm}0.07$
& $99.82{\pm}0.19$
& 94.45 \\

Lockdown$^\dagger$~\cite{NEURIPS2023_2376f25e} & 0 & 46.29
& $58.32{\pm}15.41$
& $56.41{\pm}9.59$
& $0.78{\pm}0.91$
& --$^\dagger$
& $63.45{\pm}13.34$
& $77.56{\pm}8.33$
& 51.30$^\dagger$ \\

AlignIns~\cite{Xu_2025_CVPR} & 0 & 51.28
& $\underline{33.68}{\pm}46.83$
& $\underline{0.69}{\pm}0.39$
& $\mathbf{0.35}{\pm}0.21$
& $33.72{\pm}46.81$
& $\mathbf{0.30}{\pm}0.20$
& $\mathbf{0.25}{\pm}0.17$
& \underline{11.50} \\

MASA-P10~\cite{Xu_2025_WACV} & 10 & \textbf{52.47}
& $97.29{\pm}2.27$
& $88.39{\pm}10.19$
& $98.33{\pm}0.77$
& $\mathbf{0.32}{\pm}0.32$
& $96.95{\pm}2.60$
& $99.53{\pm}0.27$
& 80.14 \\

\rowcolor{surablue}
SURA (ours) & 10 & \underline{52.43}
& $\mathbf{0.38}{\pm}0.39$
& $\mathbf{0.25}{\pm}0.09$
& $\underline{0.51}{\pm}0.35$
& $\underline{0.35}{\pm}0.21$
& $\underline{0.31}{\pm}0.13$
& $\underline{1.36}{\pm}1.77$
& \textbf{0.53} \\

\hline
\end{tabular}

\parbox{\textwidth}{\small
$^\dagger$ Scaling is omitted for Lockdown because its released
implementation does not apply the Scaling operation in our setting.
The reported mean therefore averages the remaining five attacks.}

\end{table*}

\subsection{State-Restored Response Measurement}
\label{ssec:srfp}

After $K$ train-mode probe steps, client $i$ reaches
$(\boldsymbol{\theta}_{i,K},\boldsymbol{b}_{i,K})$, so endpoint
readout evaluates different clients under different post-probe BN
states. As shown in Algorithm~\ref{alg:srfp}, SURA instead retains
$\boldsymbol{\theta}_{i,K}$ while restoring the shared pre-probe BN
state $\boldsymbol{b}^t$ only for response measurement.

To characterize what this restoration removes, let us define
\[
\ell_i^{x,y}=
\mathcal{L}_{\mathrm{p}}^{\mathrm{eval}}
(\boldsymbol{\theta}_{i,x},\boldsymbol{b}_{i,y}),
\]
where $x,y\in\{0,K\}$ index the initial or final parameter and BN
states. Then we rewrite the endpoint response
$r_{i,K}^{\mathrm{end}} := \ell_i^{K,K}-\ell_i^{0,0}$
can be decomposed as
\begin{equation}
r_{i,K}^{\mathrm{end}}
=
r_{i,K}^{\mathrm{par}}
+r_{i,K}^{\mathrm{buf}}
+r_{i,K}^{\mathrm{int}},
\label{eq:endpoint_response}
\end{equation}
The three components are defined as
\begin{align}
r_{i,K}^{\mathrm{par}}
    &=\ell_i^{K,0}-\ell_i^{0,0}, \\
r_{i,K}^{\mathrm{buf}}
    &=\ell_i^{0,K}-\ell_i^{0,0}, \\
r_{i,K}^{\mathrm{int}}
    &=\ell_i^{K,K}-\ell_i^{K,0}
      -\ell_i^{0,K}+\ell_i^{0,0}.
\end{align}
Since SURA retains $\boldsymbol{\theta}_{i,K}$ and restores the pre-probe BN state, i.e., 
$\boldsymbol{b}_{i,0}=\boldsymbol{b}^t$ for all clients $i$, 
it has the state-restored (SR) response of
\begin{equation}
\begin{aligned}
r_{i,K}^{\mathrm{SR}}
&=
\mathcal{L}_{\mathrm{p}}^{\mathrm{eval}}
(\boldsymbol{\theta}_{i,K},\boldsymbol{b}^t)
-
\mathcal{L}_{\mathrm{p}}^{\mathrm{eval}}
(\boldsymbol{\theta}_{i,0},\boldsymbol{b}^t),
\end{aligned}
\label{eq:srfp_response}
\end{equation}
which equals $r_{i,K}^{\mathrm{par}}$. Thus, the restored readout isolates the parameter-induced response by removing BN-buffer drift $r_{i,K}^{\mathrm{buf}}$ and the interaction between parameter and BN-state changes $r_{i,K}^{\mathrm{int}}$. All client responses are therefore measured under the same BN reference without altering the probing trajectory.

\subsection{Client Screening and Aggregation}
\label{ssec:scoring}

Let $v_i^t=r_{i,K}^{\mathrm{SR}}$. We normalize them across participating
clients using the median $m^t$ and median absolute deviation (MAD)
$d^t$:
\begin{equation}
m^t=\operatorname{median}_{j\in\mathcal{S}_t}v_j^t,
\quad
d^t=\operatorname{median}_{j\in\mathcal{S}_t}|v_j^t-m^t|.
\label{eq:score_normalization}
\end{equation}
We use the Gaussian-consistent scale
$s^t=1.4826 \cdot d^t$~\cite{Rousseeuw01121993}, falling back to the
population standard deviation when the MAD scale is numerically zero.
The upper-tail score is

\begin{equation}
z_i^t=
\begin{cases}
\max\left\{0,\frac{v_i^t-m^t}{s^t}\right\}, 
    & \text{if }s^t>0,\\
0,  & \text{otherwise}.
\end{cases}
\label{eq:srfp_score}
\end{equation}
Larger $z_i^t$ indicates stronger deviation from the round consensus. The threshold $\tau_{\mathrm{SR}}\ge 0$ is calibrated once offline as the $(1-q)$-quantile of scores from benign-only development runs and held fixed throughout. Client labels enter only through this one-time calibration and are never available to the server at deployment.

Accepted clients are
\begin{equation}
\mathcal{B}_t=
\{i\in\mathcal{S}_t\mid z_i^t\leq\tau_{\mathrm{SR}}\}.
\label{eq:accepted_set}
\end{equation}
Only their original updates are aggregated:
\begin{equation}
\boldsymbol{w}^{t+1}
=
\boldsymbol{w}^t+
\frac{1}{|\mathcal{B}_t|}
\sum_{i\in\mathcal{B}_t}\boldsymbol{\delta}_i^t.
\label{eq:aggregation}
\end{equation}
Since $\tau_{\mathrm{SR}}\geq0$, all clients at or below the median
are retained, ensuring $\mathcal{B}_t\neq\emptyset$.

\section{EXPERIMENTS}
\label{sec:experiments}

\subsection{Experimental Setup}
\label{ssec:experimental_setup}

\noindent\textbf{Setting.}
We benchmark on 
CIFAR-100~\cite{krizhevsky2009cifar} with
VGG16-BN~\cite{DBLP:journals/corr/SimonyanZ14a}. Client data are partitioned using a Dirichlet distribution
with concentration $\alpha=0.1$~\cite{Hsu2019}.
Training lasts 100 rounds, with all 20 clients participating in every
round and four clients acting maliciously. Each client performs two
local epochs of stochastic gradient descent (SGD) with batch size 64
and weight decay $10^{-4}$. The local learning rate starts at $0.1$
and is multiplied by $0.99$ after each round.

\noindent\textbf{Attacks and baselines.}
We consider six attacks of BadNets~\cite{8685687}, distributed
backdoor attacks (DBA)~\cite{Xie2020DBA},
Neurotoxin~\cite{pmlr-v162-zhang22w}, model-replacement
Scaling~\cite{pmlr-v108-bagdasaryan20a}, projected gradient descent
(PGD) poisoning~\cite{NEURIPS2020_b8ffa41d}, and A Little Is Enough
(LIE)~\cite{NEURIPS2019_ec1c5914}. Where applicable, malicious clients
poison 50\% of their local samples. We compare SURA with federated averaging
(FedAvg)~\cite{pmlr-v54-mcmahan17a}, the oracle FedAvg that excludes
known malicious clients, and other defenses shown in
Table~\ref{tab:main_results_vgg}.

\noindent\textbf{SURA configuration.}
SURA uses the same fixed 10-image clean proxy as all 10-image
comparison settings. We use $K=12$, ascent step size $\eta=0.001$, fusion weight
$\lambda=0.9$, momentum $\mu=0.9$, gradient norm limit
$c_{\mathrm{g}}=20$, and threshold $\tau_{\mathrm{SR}}=2.03$.
The threshold is calibrated once on development data with tail
probability $q=0.02$. Development and evaluation runs are disjoint, and the threshold
remains fixed during evaluation.

\noindent\textbf{Metrics.}
We measure clean main-task accuracy (MA) and target-label attack success rate (ASR) on triggered inputs. Malicious recall is the fraction of malicious clients rejected, and benign false-positive rate (FPR) is the fraction of benign clients incorrectly rejected. Unless otherwise stated, ASR is measured at the final round (100), and mean ASR weights the six attacks equally. Screening operates throughout training, so its decisions affect both aggregation and subsequent rounds.

\subsection{Overall Defense Performance}
\label{ssec:performance}

Table~\ref{tab:main_results_vgg} compares SURA with FedAvg, an oracle FedAvg excluding the known malicious clients, and prior defenses on CIFAR-100 with VGG16-BN, grouped by server-image budget.
Only the MASA variants, AlignIns, and SURA maintain MA close to FedAvg, while the other defenses incur substantial clean-accuracy loss.
Among defenses using at most ten server images, SURA attains the lowest mean ASR and is the only one that holds all six attacks near the oracle level.
AlignIns is the strongest competitor here, but its protection is uneven: it fails against BadNets and Scaling, where its ASR varies widely across seeds and, on some runs, it collapses.
MASA-P500 is the only prior defense that reaches mean ASRs comparable to those of the oracle FedAvg, but consumes 50$\times$ more clean server data. 
SURA matches it with only ten images, at 95.2\% malicious recall and a benign FPR of 1.65\%.

\subsection{Design Analysis}
\label{ssec:ablation}

\begin{figure}[!t]
\centering
\includegraphics[width=\columnwidth]{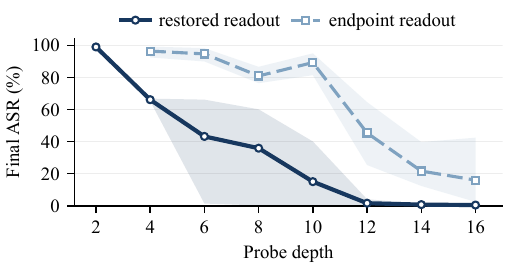}\par
\caption{Probe depth vs.\ readout on CIFAR-100/VGG16-BN using ten
proxy images. Curves average three attacks, and shading shows the
seed-wise min--max range.}
\label{fig:probe_depth}
\end{figure}

\noindent\textbf{Repeated probing.}
We vary repeated clean-loss ascent steps $K$ using the same ten-image
proxy (Fig.~\ref{fig:probe_depth}). Across both endpoint and restored
readouts, deeper probing substantially improves attack suppression
overall. From $K=4$ to $K=16$, mean ASR drops from 96.49\% to
16.02\% with endpoint readout and from 66.18\% to 0.49\% with
restored readout. Although the curves are not monotonic at every
intermediate depth, the overall trend shows that repeated probing
itself improves detection. With restored readout, SURA reaches
approximately 1.64\% ASR at $K=12$ in this three-attack evaluation.
These results support repeated probing as an effective design under
scarce proxy data.

\noindent\textbf{Response readout.}
Fig.~\ref{fig:probe_depth} also shows consistently lower ASR with
restored readout at matched probe depths. To isolate this effect at
fixed depth, we compare accumulated, endpoint, and restored endpoint
readouts at $K=12$ under matched benign-FPR calibration.
Table~\ref{tab:readout_ablation} shows that endpoint readout
substantially reduces mean ASR and BN restoration yields a further
reduction. Both reductions exceed twice the seed-level standard
errors. The restored row uses matched calibration rather than the
deployed point in Table~\ref{tab:main_results_vgg}. Together with
Fig.~\ref{fig:probe_depth}, these results support treating probe depth
and response readout as separate design choices.

\begin{table}[!t]
\begingroup
\centering
\caption{Matched readout ablation at $K=12$ on CIFAR-100 with
VGG16-BN under matched benign-FPR calibration.}
\label{tab:readout_ablation}
\small
\setlength{\tabcolsep}{4pt}
\begin{tabular}{lccc}
\hline
Readout & ASR$\downarrow$ & MA$\uparrow$ & Recall$\uparrow$ \\
\hline
Accumulated & 81.70 & 53.07 & 42.03 \\
Endpoint & 29.00 & 52.73 & 84.09 \\
Endpoint + BN restore & 0.83 & 52.54 & 93.77 \\
\hline
\end{tabular}
\par
\endgroup
\end{table}

\begin{table}[!b]
\begingroup
\centering
\caption{Proxy sensitivity, CIFAR-100/VGG16-BN, settings fixed. Draw~1 is used in all other experiments; entries are percentages.}
\label{tab:proxy_sensitivity}
\small
\begin{tabular}{lcccc}
\hline
Proxy & ASR$\downarrow$ & MA$\uparrow$ & Recall$\uparrow$ & FPR$\downarrow$ \\
\hline
Draw 1 & 0.462 & 52.40 & 95.94 & 1.60 \\
Draw 2 & 0.437 & 52.63 & 94.76 & 1.41 \\
Draw 3 & 0.968 & 52.55 & 91.97 & 2.05 \\
Draw 4 & 0.449 & 52.56 & 96.15 & 1.51 \\
\hline
\end{tabular}
\par
\endgroup
\end{table}

\noindent\textbf{Proxy sensitivity.} Since the ten proxy images are sampled once and reused throughout training, we test whether detection depends on their selection. We repeat the evaluation with three independently drawn ten-image sets across six attacks and three common training seeds. All defense settings remain fixed, including the threshold, which is not recalibrated. Across the four draws, mean ASR ranges from 0.437\% to 0.968\%, while MA remains between 52.40\% and 52.63\% (Table~\ref{tab:proxy_sensitivity}). Thus, SURA maintains stable performance across the tested proxy selections without threshold adjustment.

\subsection{Data and Computational Efficiency}
\label{ssec:efficiency}

Table~\ref{tab:efficiency} tests whether more unlearning iterations
or proxy data alone explain SURA's performance. Matching SURA's
ascent-step budget on the same ten-image proxy improves MASA-P10,
but remains far less effective than SURA. Enlarging the MASA proxy
also improves attack suppression, yet does not reproduce SURA's
low-data performance. Neither control alone explains SURA's gains.

Compared with MASA-P500, SURA uses 3.33$\times$ fewer ascent steps
and 20.8$\times$ fewer proxy examples in backward passes, with lower
isolated round time. BN restoration itself adds no gradient step.
Thus, SURA approaches large-proxy robustness with substantially lower
proxy and gradient requirements.

\begin{figure}[!t]
\centering
\includegraphics[width=\columnwidth]{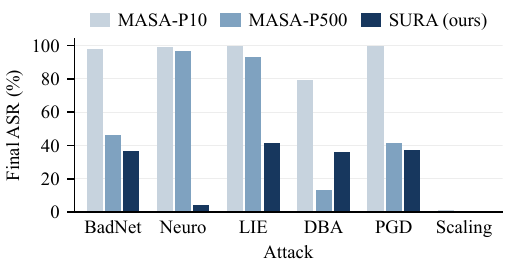}\par
\caption{Attack-wise ASR, CIFAR-10/ResNet18, $\alpha{=}0.5$. Proxy size: 10 (SURA, MASA-P10) vs.\ 500 (MASA-P500).}
\label{fig:resnet_asr}
\end{figure}

\begin{table}[!t]
\centering
\caption{Data and computation, CIFAR-100/VGG16-BN. Steps and Bwd.~ex.\ are
server ascent steps and proxy examples in backward passes, per round;
Time is relative to FedAvg.}
\label{tab:efficiency}
\small
\setlength{\tabcolsep}{2.6pt}
\begin{tabular}{lrrrrr}
\hline
Method & Images & Steps & Bwd. ex. & ASR & Time \\
\hline
FedAvg & -- & 0 & 0 & 99.32 & $1.000\times$ \\
MASA-P10 & 10 & 100 & 1,000 & 80.14 & $1.046\times$ \\
MASA-P10 (12 ep.) & 10 & 240 & 2,400 & 61.75 & -- \\
MASA-P120 & 120 & 200 & 12,000 & 25.57 & -- \\
SURA & 10 & 240 & 2,400 & 0.53 & $1.157\times$ \\
MASA-P500 & 500 & 800 & 50,000 & 0.35 & $1.479\times$ \\
\hline
\end{tabular}
\end{table}

\subsection{Generalization}
\label{ssec:cifar10}

We further benchmark SURA on CIFAR-10~\cite{krizhevsky2009cifar}
with an 18-layer residual network (ResNet18)~\cite{He_2016_CVPR} under non-IID $\alpha=0.5$,
using the CIFAR-100/VGG16-BN configuration unchanged without
ResNet-specific recalibration. Across six attacks, SURA achieves
26.13\% mean ASR, compared with 79.55\% for MASA-P10 and
48.65\% for MASA-P500 (Fig.~\ref{fig:resnet_asr}). These correspond
to reductions of 53.42 and 22.52 percentage points, respectively.
MA remains comparable at 87.96\% for SURA, 87.70\% for MASA-P10,
and 87.94\% for MASA-P500. This zero-shot transfer outperforms
MASA-P500 with only ten proxy images, indicating that proxy size
alone does not explain the performance difference.

\subsection{Adaptive Attacks}
\label{ssec:adaptive}

We evaluate defense-aware attacks~\cite{pmlr-v108-bagdasaryan20a} at $K=12$ by varying proxy knowledge and the allowed update modifications. Scale-only adaptation changes the update magnitude at a fixed direction; direction-only adaptation changes the direction at a fixed magnitude. Joint adaptation optimizes both.

When the exact server proxy is unavailable to the adversary, scale-only, direction-only, and joint adaptation yield ASRs of 0.37\%, 0.33\%, and 0.37\%, respectively. We then disclose the exact proxy: direction-only adaptation yields 0.73\% ASR, but joint direction-and-scale adaptation reaches 94.91\%. This contrast identifies exact-proxy disclosure as an important operating limitation of SURA.

\section{Conclusion}

We introduced SURA, which separates repeated clean-loss probing from
response measurement for data-efficient FL backdoor detection. SURA
strengthens client responses through repeated probing and measures
them under a shared pre-probe BN state to avoid client-specific BN
drift. Across six attacks, SURA achieves low ASR with ten clean server
images while preserving main-task accuracy and approaching large-proxy
robustness. These results suggest that jointly controlling probe depth
and readout state is an effective principle for low-data backdoor
screening.

\clearpage

\section{Compliance with Ethical Standards}
This study consists of numerical experiments using the publicly
available CIFAR-10 and CIFAR-100 benchmark datasets in simulated
federated learning settings. It does not involve human or animal
subject research, and no ethical approval was required.

\bibliographystyle{IEEEtran}
\bibliography{refs}

\begin{thebibliography}{10}
\providecommand{\url}[1]{#1}
\csname url@samestyle\endcsname
\providecommand{\newblock}{\relax}
\providecommand{\bibinfo}[2]{#2}
\providecommand{\BIBentrySTDinterwordspacing}{\spaceskip=0pt\relax}
\providecommand{\BIBentryALTinterwordstretchfactor}{4}
\providecommand{\BIBentryALTinterwordspacing}{\spaceskip=\fontdimen2\font plus
\BIBentryALTinterwordstretchfactor\fontdimen3\font minus \fontdimen4\font\relax}
\providecommand{\BIBforeignlanguage}[2]{{%
\expandafter\ifx\csname l@#1\endcsname\relax
\typeout{** WARNING: IEEEtran.bst: No hyphenation pattern has been}%
\typeout{** loaded for the language `#1'. Using the pattern for}%
\typeout{** the default language instead.}%
\else
\language=\csname l@#1\endcsname
\fi
#2}}
\providecommand{\BIBdecl}{\relax}
\BIBdecl

\bibitem{pmlr-v54-mcmahan17a}
B.~McMahan, E.~Moore, D.~Ramage, S.~Hampson, and B.~A. y~Arcas, ``Communication-efficient learning of deep networks from decentralized data,'' in \emph{Proc. AISTATS}, vol.~54, 2017, pp. 1273--1282.

\bibitem{pmlr-v108-bagdasaryan20a}
E.~Bagdasaryan, A.~Veit, Y.~Hua, D.~Estrin, and V.~Shmatikov, ``How to backdoor federated learning,'' in \emph{Proc. AISTATS}, vol. 108, 2020, pp. 2938--2948.

\bibitem{8685687}
T.~Gu, K.~Liu, B.~Dolan-Gavitt, and S.~Garg, ``Badnets: Evaluating backdooring attacks on deep neural networks,'' \emph{IEEE Access}, vol.~7, pp. 47\,230--47\,244, 2019.

\bibitem{NIPS2017_f4b9ec30}
P.~Blanchard, E.~M. {El Mhamdi}, R.~Guerraoui, and J.~Stainer, ``Machine learning with adversaries: {Byzantine} tolerant gradient descent,'' in \emph{Advances in Neural Information Processing Systems}, vol.~30, 2017, pp. 119--129.

\bibitem{259745}
C.~Fung, C.~J.~M. Yoon, and I.~Beschastnikh, ``The limitations of federated learning in {Sybil} settings,'' in \emph{Proc. RAID}, 2020, pp. 301--316.

\bibitem{280048}
T.~D. Nguyen, P.~Rieger, H.~Chen, H.~Yalame, H.~M{\"o}llering, H.~Fereidooni, S.~Marchal, M.~Miettinen, A.~Mirhoseini, S.~Zeitouni, F.~Koushanfar, A.-R. Sadeghi, and T.~Schneider, ``{FLAME}: Taming backdoors in federated learning,'' in \emph{Proc. USENIX Security}, 2022, pp. 1415--1432.

\bibitem{9721118}
K.~Pillutla, S.~M. Kakade, and Z.~Harchaoui, ``Robust aggregation for federated learning,'' \emph{IEEE Transactions on Signal Processing}, vol.~70, pp. 1142--1154, 2022.

\bibitem{Huang_2023_ICCV}
S.~Huang, Y.~Li, C.~Chen, L.~Shi, and Y.~Gao, ``Multi-metrics adaptively identifies backdoors in federated learning,'' in \emph{Proceedings of the IEEE/CVF International Conference on Computer Vision (ICCV)}, October 2023, pp. 4652--4662.

\bibitem{Xu_2025_WACV}
J.~Xu, Z.~Zhang, and R.~Hu, ``Identify backdoored model in federated learning via individual unlearning,'' in \emph{Proceedings of the Winter Conference on Applications of Computer Vision (WACV)}, February 2025, pp. 7949--7958.

\bibitem{pmlr-v37-ioffe15}
S.~Ioffe and C.~Szegedy, ``Batch normalization: Accelerating deep network training by reducing internal covariate shift,'' in \emph{Proc. ICML}, vol.~37, 2015, pp. 448--456.

\bibitem{DBLP:conf/ndss/CaoF0G21}
X.~Cao, M.~Fang, J.~Liu, and N.~Z. Gong, ``Fltrust: Byzantine-robust federated learning via trust bootstrapping,'' in \emph{28th Annual Network and Distributed System Security Symposium, {NDSS} 2021, virtually, February 21-25, 2021}.\hskip 1em plus 0.5em minus 0.4em\relax The Internet Society, 2021.

\bibitem{NEURIPS2023_2376f25e}
T.~Huang, S.~Hu, K.-H. Chow, F.~Ilhan, S.~Tekin, and L.~Liu, ``Lockdown: Backdoor defense for federated learning with isolated subspace training,'' in \emph{Advances in Neural Information Processing Systems}, A.~Oh, T.~Naumann, A.~Globerson, K.~Saenko, M.~Hardt, and S.~Levine, Eds., vol.~36.\hskip 1em plus 0.5em minus 0.4em\relax Curran Associates, Inc., 2023, pp. 10\,876--10\,896.

\bibitem{Xu_2025_CVPR}
J.~Xu, Z.~Zhang, and R.~Hu, ``Detecting backdoor attacks in federated learning via direction alignment inspection,'' in \emph{Proceedings of the IEEE/CVF Conference on Computer Vision and Pattern Recognition (CVPR)}, June 2025, pp. 20\,654--20\,664.

\bibitem{Rousseeuw01121993}
P.~J. Rousseeuw and C.~Croux, ``Alternatives to the median absolute deviation,'' \emph{Journal of the American Statistical Association}, vol.~88, no. 424, pp. 1273--1283, 1993.

\bibitem{krizhevsky2009cifar}
A.~Krizhevsky, ``Learning multiple layers of features from tiny images,'' University of Toronto, Tech. Rep., 2009.

\bibitem{DBLP:journals/corr/SimonyanZ14a}
\BIBentryALTinterwordspacing
K.~Simonyan and A.~Zisserman, ``Very deep convolutional networks for large-scale image recognition,'' in \emph{3rd International Conference on Learning Representations, {ICLR} 2015, San Diego, CA, USA, May 7-9, 2015, Conference Track Proceedings}, Y.~Bengio and Y.~LeCun, Eds., 2015. [Online]. Available: \url{http://arxiv.org/abs/1409.1556}
\BIBentrySTDinterwordspacing

\bibitem{Hsu2019}
\BIBentryALTinterwordspacing
H.~Hsu, H.~Qi, and M.~Brown, ``Measuring the effects of non-identical data distribution for federated visual classification,'' 2019. [Online]. Available: \url{https://arxiv.org/abs/1909.06335}
\BIBentrySTDinterwordspacing

\bibitem{Xie2020DBA}
C.~Xie, K.~Huang, P.-Y. Chen, and B.~Li, ``Dba: Distributed backdoor attacks against federated learning,'' in \emph{International Conference on Learning Representations}, 2020.

\bibitem{pmlr-v162-zhang22w}
Z.~Zhang, A.~Panda, L.~Song, Y.~Yang, M.~Mahoney, P.~Mittal, R.~Kannan, and J.~Gonzalez, ``{Neurotoxin}: Durable backdoors in federated learning,'' in \emph{Proc. ICML}, vol. 162, 2022, pp. 26\,429--26\,446.

\bibitem{NEURIPS2020_b8ffa41d}
H.~Wang, K.~Sreenivasan, S.~Rajput, H.~Vishwakarma, S.~Agarwal, J.-y. Sohn, K.~Lee, and D.~Papailiopoulos, ``Attack of the tails: Yes, you really can backdoor federated learning,'' in \emph{Advances in Neural Information Processing Systems}, H.~Larochelle, M.~Ranzato, R.~Hadsell, M.~Balcan, and H.~Lin, Eds., vol.~33.\hskip 1em plus 0.5em minus 0.4em\relax Curran Associates, Inc., 2020, pp. 16\,070--16\,084.

\bibitem{NEURIPS2019_ec1c5914}
G.~Baruch, M.~Baruch, and Y.~Goldberg, ``A little is enough: Circumventing defenses for distributed learning,'' in \emph{Advances in Neural Information Processing Systems}, H.~Wallach, H.~Larochelle, A.~Beygelzimer, F.~d\textquotesingle Alch\'{e}-Buc, E.~Fox, and R.~Garnett, Eds., vol.~32.\hskip 1em plus 0.5em minus 0.4em\relax Curran Associates, Inc., 2019.

\bibitem{He_2016_CVPR}
K.~He, X.~Zhang, S.~Ren, and J.~Sun, ``Deep residual learning for image recognition,'' in \emph{Proceedings of the IEEE Conference on Computer Vision and Pattern Recognition (CVPR)}, June 2016.

\end{thebibliography}

\end{document}